\documentclass[10pt]{article}
\usepackage[letterpaper]{geometry}
\usepackage{hicss51}
\usepackage{times}
\usepackage[none]{hyphenat}
\usepackage{url}
\usepackage{latexsym}
\usepackage{indentfirst}
\usepackage{graphicx}
\graphicspath{{images/}}

\usepackage{lipsum}
\usepackage{mathptmx}
\usepackage{subfigure}

\usepackage{amssymb}

\usepackage{flushend}

\usepackage{multirow}
\usepackage{hhline}
\usepackage{adjustbox}
\usepackage{color}
\usepackage[linesnumbered,lined,commentsnumbered]{algorithm2e}
\usepackage{algpseudocode}
\usepackage{pifont}
\usepackage{xspace}

\usepackage{breakurl}
\usepackage[breaklinks]{hyperref}
\usepackage{amsmath}

\usepackage{enumitem}
\usepackage{ragged2e}
\usepackage{xcolor,colortbl}
\colorlet{red}{black}

\title{Route Packing: Geospatially-Accurate Visualization of Route Networks}

\author{
Jieqiong Zhao\\
Purdue University\\ \underline{\smash{zhao413@purdue.edu}} \\
\\
Shehzad Afzal\\
King Abdullah University \\of Science and Technology\\ \underline{\smash{shehzad.afzal@kaust.edu.sa}}
\And
Morteza Karimzadeh\\
University of Colorado Boulder\\ \underline{\smash{karimzadeh@colorado.edu}}\\
\\
Guizhen Wang\\
Purdue University\\ \underline{\smash{wang1908@purdue.edu}}
\And
Hanye Xu\\
Purdue University\\ \underline{\smash{hanye.vera.xu@gmail.com}}\\
\\
Niklas Elmqvist\\
University of Maryland College Park\\ \underline{\smash{elm@umd.edu}}
\And
Abish Malik\\
Davista Technologies\\ \underline{\smash{amalik@davistatechnologies.com}}\\
\\
David S. Ebert\\
Purdue University\\ \underline{\smash{ebertd@purdue.edu}}
}

\date{}

\begin{document}
\maketitle

\begin{abstract}
  We present \textit{route packing}, a novel (geo)visualization technique for displaying several routes simultaneously on a
  geographic map while preserving the geospatial layout, identity, directionality, and
  volume of individual routes.
  The technique collects variable-width route lines side by side while minimizing crossings,
  encodes them with categorical colors, and decorates them with glyphs
  to show their directions.
  Furthermore, nodes representing sources and sinks 
  use glyphs to indicate whether routes stop at the node or
  merely pass through it.
  We conducted a crowd-sourced user study investigating route tracing
  performance with road networks visualized using our route packing
  technique.
  Our findings highlight the visual parameters under which the
  technique yields optimal performance.
\end{abstract}
\vspace{-2em}
\begin{figure*}[thb]
  \centering
  \includegraphics[width=0.8\textwidth]{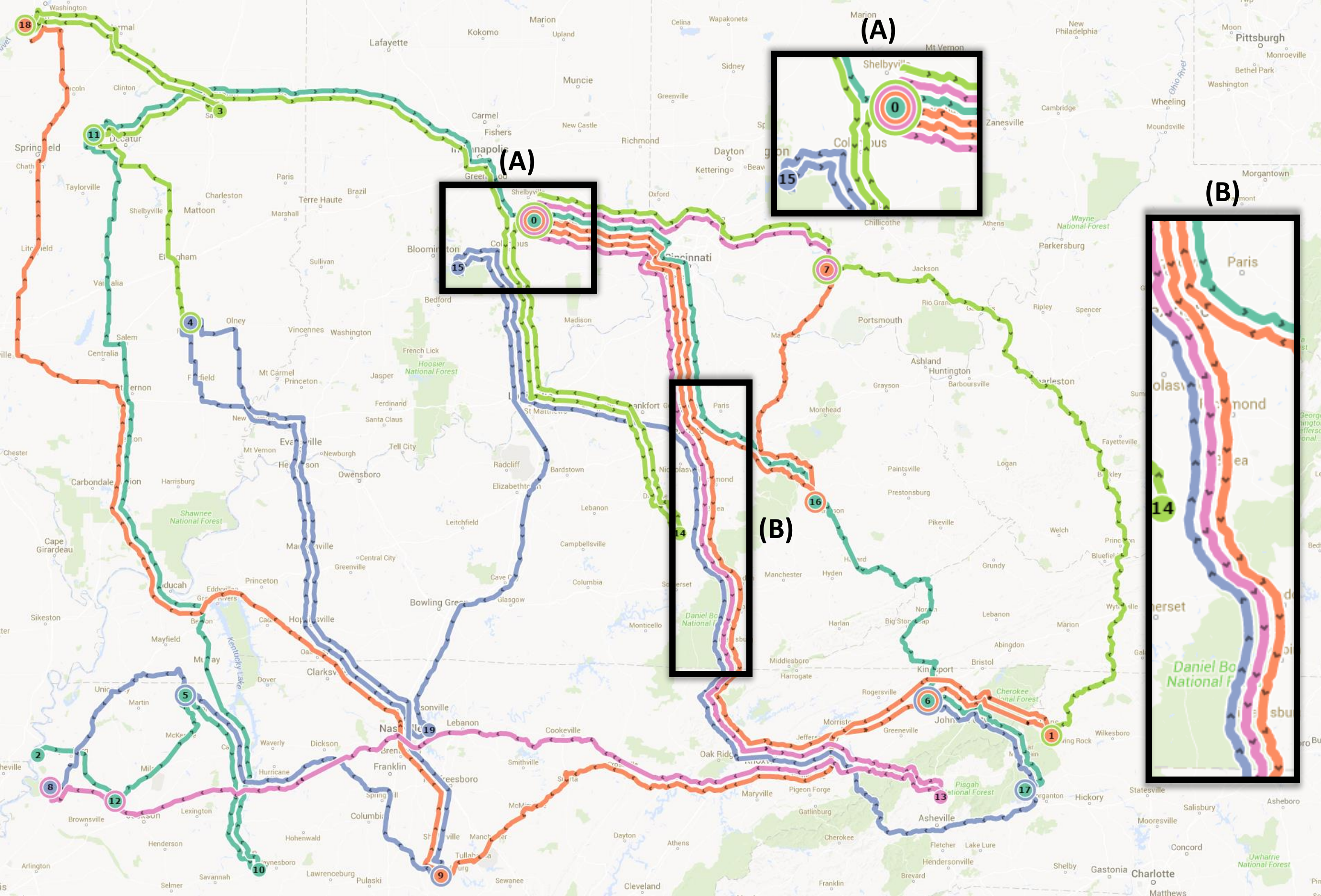}
  \caption{
    Route packing for a supply-chain network consisting of
    multiple individual routes (distinguished by categorical color
    encoding).
    Routes that share the same road for part or all of their stretch
    are packed rather than merged into a single line, preserving their
    identity. Glyphs on routes and on nodes indicate the direction of the route and whether they pass or stop at a specific node.}
     \vspace{-0.5cm}    
  \label{fig:teaser}
\end{figure*}

\section{Introduction}
Visualizing several routes on a geographic map while preserving the geospatial layout of the routes is a challenging task. Routes may overlap or cross each other, merge and depart again, share the same edge but have opposite directions, and have different semantics, speeds or volumes. Many state-of-the-art techniques
tend to use visual or data grouping techniques---such as edge
bundles~\cite{Holten2006, Holten2009}, flow map
layouts~\cite{Phan2005}, and metro maps~\cite{Shen2007, Wu2012}--- that
not only cluster entities into groups but also simplify their exact
movement paths into schematic maps. 
Based on the same principles as Harry Beck's tube map from
1931~\cite{Degani2013}, these techniques reduce complex paths into
straight lines and organize haphazardly scattered nodes onto a regular
grid to facilitate comprehension and legibility.

But what if we are truly interested in the exact geospatial position and layout
of a particular path?
Consider a supply-chain network of multiple trucks transporting goods
from a distribution center to a set of receiving sites on a daily
basis, such as for a fast food restaurant chain.
What if we need to be able to distinguish individual routes, even when
they temporarily merge together on the same physical road, only to
separate again?
What if we need to be able to follow a route that passes several nodes
(i.e., restaurants) without stopping before reaching its destination?
And what if we also need to preserve the temporal order and volume of
goods being transported on each individual route? What if there is a high risk of flooding (or other kinds of risks) in the covered area, and the planners need to route the trucks while overlaying maps with flood (or other kinds of risk) maps to mitigate the delivery risk? Distorting the nodes and routes into schematic maps in such scenarios at best slows the human decision-maker down, and may lead to confusion or poor planning. These additional constraints conflict with the visual and data
aggregation techniques reviewed above, highlighting that current techniques are ill-equipped to handle
situations in which the geospatial layout provides indispensable geographic context. 

Compared to schematic maps (e.g., metro maps~\cite{Shen2007, Wu2012}),
geospatially-accurate maps (henceforth referred to as geo-accurate) preserve the exact geographic location of entity paths and are thus more understandable to humans.
For instance, London Underground publishes a geo-accurate metro map~\cite{Brownlee2015}, particularly
after Guo~\cite{Guo2011} estimated that up to 30\% of London travelers
take wrong routes with the schematic metro map.
Geo-accurate maps can be particularly important for tasks where location is critical to decision-making.
For example, supply-chain companies monitor road conditions of
specific locations when a hurricane may block their delivery
routes~\cite{Ko2014}.
Similarly, Google Maps provides traffic and accident locations on
the map (for a single route) to help users select a fast and safe route.
Although geo-accurate maps are important in
decision-making, research on how such maps can be best visualized remains limited, especially on visualizing several routes. 

In this paper, we present a novel technique for geo-accurate visualization of routes called \textit{route packing}
(Figure~\ref{fig:teaser}) that groups multiple routes sharing the same
geographic space side-by-side without bundling them (while minimizing crossings), thereby
preserving the identity of each individual route.
While similar to metro maps~\cite{Wu2012}, route packing does not
simplify routes into straight or diagonal lines, but instead preserves
their length, direction, and geospatial layout, making it useful for logistics planning and applications where geographic context is important.
The route packing algorithm is sufficiently flexible to encode volume (such as amount of goods, traffic, or cost) \textcolor{red}{into} the width of the graphical line representing each route. 
To indicate whether a route stops \textcolor{red}{at} a specific node or merely passes it by, we use glyph decorations on nodes. 
\textcolor{red}{Route packing is appropriate for visualizing 10-20 routes simultaneously (typical of supply-chain planning scenarios), enabling decision makers to compare the routes while preserving geographic context.
Figure~\ref{fig:teaser} presents an exemplar supply chain network from a fast food company, with five trucks delivering products to 20 stores, where each truck stops at 5--9 stores.}

We also evaluate our route packing technique using
a crowd-sourced user study involving more than 100 participants.
Since no directly-comparable visualization technique exists in the
literature, we opted to measure user performance using variations in
route packing design. We applied the design choices proposed by Holten et
al.~\cite{Holten2011, Holten2009b} for visualizing directed node-link
diagrams to route packing, comparing visual perception for arrows,
tapered routes, and transparency. Our findings show that redundant encoding using both arrows and
tapered lines resulted in the best performance for tracing routes in
the network.

\section{Related Work}\label{sec:background}
Route networks can be visualized in three major ways~\cite{Steiger2014}:
When the geospatial position of vertices and edges is important, geo-accurate (or geo-referenced) networks are required, which can further be organized into traditional space-preserving vs.\
distortion-based representations. On the other hand, sometimes only the connectivity of the network is needed, in which general graph visualization is sufficient. 
Metro maps are a hybrid approach in which the network is laid out to maximize legibility, yet maintain some relation to the geographic position of the vertices and edges. 
Below we review these topics in more detail.


To show multiple routes simultaneously in \textbf{geo-accurate maps}, many techniques have utilized entity clustering algorithms with flowlines, including Phan et al.'s
flow map layout~\cite{Phan2005}, Andrienko and Andrienko's silhouette graphs~\cite{Andrienko2004}, Hadlak et al.'s attributed hierarchical structures that change over time~\cite{Hadlak2010}, and Bouvier and Oates's staining with flow
arrows~\cite{Bouvier2008}. Cornel et al. recently presented composite flow maps ~\cite{cornel2016}, which extract flows based on the structure of route shapes but then abstract and combine flows as ribbons. Our route packing technique stays faithful to the geospatial layout of the route network (helping users perceive the length of routes), preserving the individual identity and direction of routes while minimizing displacement and crossings.

\textbf{Distortion-based} methods distort the geometries in geographic space to optimize legibility or visual search.
Cartograms, where area and distances in a map are distorted based on a variable such as population, are the origin such techniques~\cite{Tobler1976}, leading to work such as M{\'e}lange space-folding~\cite{Elmqvist2010}, and the generalized
transmogrification space distortion framework~\cite{Brosz2013} for keeping visibility by distorting space.

Our route packing technique does not distort the geographic space of
the map itself, but it does distort the geometric stretch of the
routes in order to avoid occlusion.
This is essentially similar to a static version of the
EdgeLens~\cite{Wong2003}, which provides an interaction to visually
separate multiple edges around a node to optimize their visibility.


Modeling a route network as a graph is possible if only the
connectivity of the network is significant. Current \textbf{graph layout}
algorithms can be used to find optimal positions for the
vertices and edges~\cite{Battista1999}.
Some graph visualization techniques have focused on preserving the overall 
topology (e.g.,~\cite{Dwyer2008}) or intelligent routing of edges~\cite{Dobkin1997,
  Dwyer2009}. However, while these efforts fulfill part of our needs in visualizing
route networks, they are not suitable for geo-accurate maps.

Similar to geospatial visualization, scale is a key challenge for graph
visualization.
Early solutions focused on clustering vertices into
groups~\cite{Eades1996}, but recent work has instead used edge
bundling~\cite{Holten2006, Holten2009} where edges are
grouped together to reduce occlusion.
Interestingly, current edge bundling approaches are reminiscent of the
flow map and flowline layouts discussed above.
However, clustering and bundling for graphs have the same drawback,
which is that the identity of individual entities is not preserved.

Nevertheless, we draw on prior graph
visualization research. Holten and van Wijk in 2009 presented an in-depth user study on edge tracing in node-link diagrams~\cite{Holten2009b}, and in 2011, extended the study to a wider set of visual representations~\cite{Holten2011}. Our user study in this paper is based on these two studies, and we also derive our visual representations for directionality from their design space exploration.

\textbf{Metro map} layouts usually use vertical, horizontal, 
and diagonal edge directions, but maintain the geographic relations
between nodes (or stations)~\cite{Steiger2014}. Wu et al.~\cite{Wu2013,
Wu2012} studied the visualization aspects of automatically
generating metro maps from a geographic map and connectivity
graph.
While this kind of layout has does not respect the geographic locations of edges, there are overarching
design ideas here that are worth adopting. For example, Shen and Ma~\cite{Shen2007} use map layouts to visualize multiple parallel paths in an adjacency matrix representation of a
graph. Similarly, Alper et al.~\cite{Alper2011} provide a linearized view of
their LineSets visualization using the familiar visual language of
metro maps. Our route packing technique builds on
these ideas; we juxtapose incident routes side by side without
bundling them together into a single unit, and use metro station
visual encodings to communicate route crossings and shared stops.



Excessive crossings in metro maps can cause visual clutter and confusion for the end-user.
The metro-line crossing minimization (MLCM) problem was proposed and formulated by Benkert et al.~\cite{benkert2007} in 2007. Bekos et al.~\cite{Bekos2008} studied the MLCM problem on the graph structure of a path or tree. 
Later, Asquith et al.~\cite{Asquith2008} applied integer linear programming to solve the problem for general graphs. N{\"{o}}llenburg~\cite{Nollenburg2010} and Mink at el. ~\cite{Fink2013,Fink2015} applied their line ordering methods on embedded graphs with an obvious relationship between vertices and edges.
Extending these design principles from the MLCM research, we adapted the prior lines placement algorithms to keep the relative order along the edges and change the order in vertices.
\section{The Route Packing Technique}\label{sec:design}
\textit{Route packing} juxtaposes overlapping routes side by side to keep the identity of each route distinguishable while staying faithful to its geographic footprint. In this section, we describe the details of our approach.  

\subsection{Data Model}\label{sec:data-model}
The data model for route network visualization consists of a
georeferenced graph $G = < V, E >$ modeling both the topology and the
geographic position of the network as well as a set of routes $R = \{
r_0, r_1, \dots, r_n \}$.
A route $r_i = < V_i, E_i >$ is a list of vertices $V_i = ( v_{i,0},
v_{i, 1}, \dots, v_{i, n})$ representing the stops along the way, as
well as a list of edges $E_i = ( e_{i,0}, e_{i, 1}, \dots, e_{i, n-1})$
representing the path between them.

\begin{figure}[htb]
  \centering
  \resizebox{\columnwidth}{!}{\includegraphics{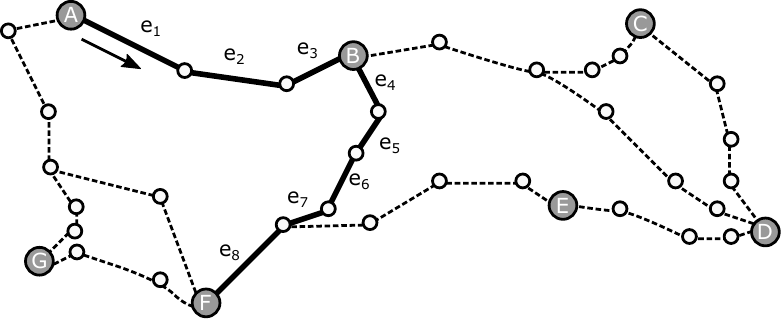}}
  \caption{Example of a route in a route network.
    The highlighted route consists of two legs---$A$ to $B$, and $B$
    to $F$---where the route stops temporarily, as well as the
    edges $e_1, \dots, e_8$ making up the actual path.}
  \label{fig:route-network}
  \vspace{-0.8em}
\end{figure}

Unlike some road networks that model only the major
cities in the network using vertices, the georeferenced graph uses
vertices with specific geographic positions to model the geometry of the road.
Figure~\ref{fig:route-network} shows an example of a small route
network with major nodes $A, B, \dots, G$, and many minor vertices
modeling waypoints along the roads.
Since the edge list in a route $r_i$ contains sufficient information
to determine its geometry, the vertex list in the
route is only used to determine which of the vertices along the route
are stopping points.
We call each consecutive pair of vertices in a route a \textit{leg}.
For example, the highlighted route in Figure~\ref{fig:route-network}
consists of the legs $(A, B)$ and $(B, F)$.

\subsection{Design Rationale}\label{sec:rationale}
Given this data model, we formalize our design requirements
for the route packing technique as follows:

\setlist{leftmargin = 5.4mm}
\begin{itemize}
\itemsep0em
\item[R1]\textbf{Preserve route edges:} All edges in a route
  network should be preserved so that they can be distinguished from
  other edges and matched to routes.
  
\item[R2]\textbf{Preserve route legs:} Legs along a route should be
  distinguishable so that endpoints can be identified.
  
\item[R3]\textbf{Convey direction:} The direction of a particular
  route should be conveyed across its entire path.
  
\item[R4] [Optional] \textbf{Convey volume:} Legs can optionally be
  associated with a weight representing volume, e.g.\ the amount of
  goods being transported.
\end{itemize}

The challenge for virtually all of these requirements is that a
typical multi-route network contains a significant amount of overlap.
For R1 (preservation of edges), the problem is that many routes will
be sharing the same geographic stretch of a road for a partial (or full) leg.
Sometimes the same stretch of road will be shared by two
different routes heading in opposite directions, which challenges R3.
For R3, routes may \textcolor{red}{pass} by certain vertices without
stopping, whereas other routes do stop.
Here, the challenge is to convey whether or not a particular route
stops at a vertex.


\begin{figure}[htb]
  \centering
  \subfigure[Occlusion of edges in route.]{
    \resizebox{0.46\columnwidth}{!}{\includegraphics{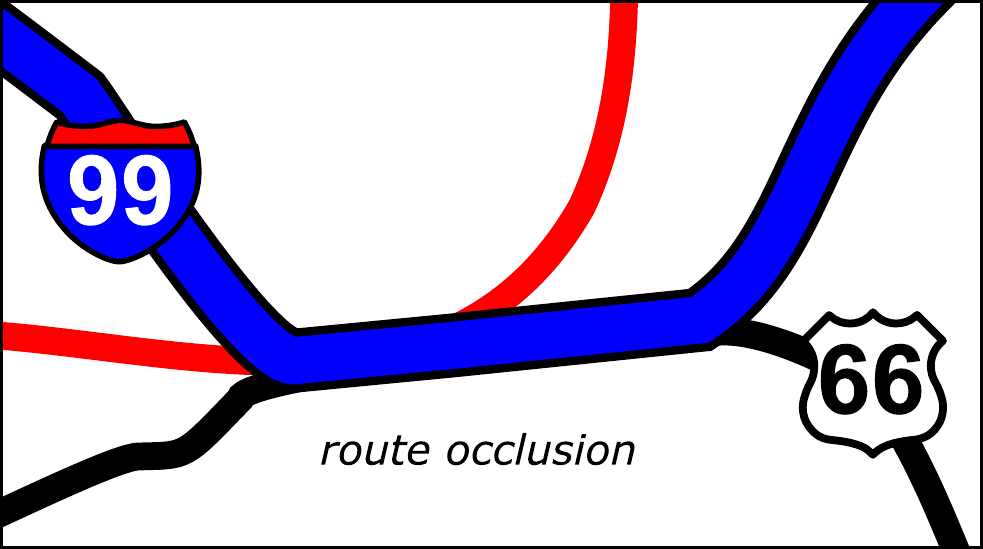}}
    \label{fig:packing-before}
  }
  \subfigure[Route packing to preserve edges.]{
    \resizebox{0.46\columnwidth}{!}{\includegraphics{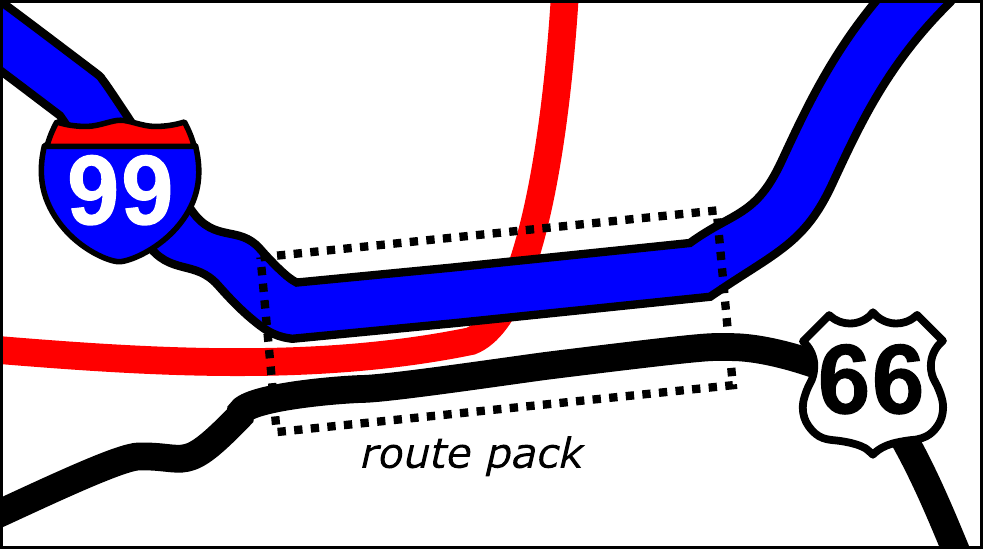}}
    \label{fig:packing-after}
  }
  \caption{Example of three routes in (a) with significant overlap in
    a shared part of the road, causing occlusion.
    In (b), route packing has been applied to separate
    edges so that they are distinguishable and identifiable.
    Edge crossings are sometimes inevitable.}
  \label{fig:packing}
	\vspace{-0.8em}
\end{figure}


\subsection{Visual Design: Route Layout}
\label{sec:route-layout}
Since different routes may share the same stretch of a road, directly visualizing different routes
can generate severe visual confusion and hinder effective recognition
of routes.
For example, in Figure~\ref{fig:packing-before}, the visual overlapping
of the three routes makes it hard for a viewer to see detailed
geographical information about the occluded edges.
Route packing shifts and packs route edges to minimize visual
overlap, resulting in the packed layout in
Figure~\ref{fig:packing-after}.
The layout algorithm consists of several steps:

\setlist{leftmargin = 4.2mm}
\begin{enumerate}
\itemsep0em
\item\textbf{Detecting overlap}: Due to the high precision
  of GPS coordinates, small data errors in GIS databases, and
  specific terrain conditions such as varying road width, route edges
  do not necessarily have to share the exact coordinates to
  exhibit overlap when rendered on the screen.
  To detect overlapping edges, we apply linear Kernel
  Density Estimation 
  (KDE)
  to build a route network skeleton that groups
  route edges that are in close geographic proximity.
  Route edges are overlapped if they share some part of the
  skeleton.
  
\item\textbf{Shifting and packing:} Once overlapping edges have been detected, we \textit{shift} them (displace them from their original position) to eliminate the overlap, and then \textit{pack} them (bring them together along the same path while respecting edge width) one by one (Figure~\ref{fig:packing-after}).
The displacement should be minimal and perpendicular to the edge direction for consistency.
When displacing one edge, another edge that was not previously overlapped may become occluded.
Therefore, these two stages---detection and shift/pack---are iterative and must be repeated until no more overlap is detected.
When two edges overlap, only one of them must be shifted and packed.
\textcolor{red}{We apply route crossing minimization to decide which edge should have priority over the others (more details in Section 3.6).}

\item\textbf{Rendering:} We use a categorical color assignment based
  on the topology of the route network skeleton (from Step 1) to render adjacent routes with different colors.
  Additionally, we use a white halo around each route line to visually separate packed routes.

\end{enumerate}

\subsection{Visual Design: Route Direction}
\begin{figure}[htb]
    \vspace{-1.5em}
  \centering
  \subfigure[Arrow glyphs]{
    \resizebox{0.30\columnwidth}{!}{\includegraphics{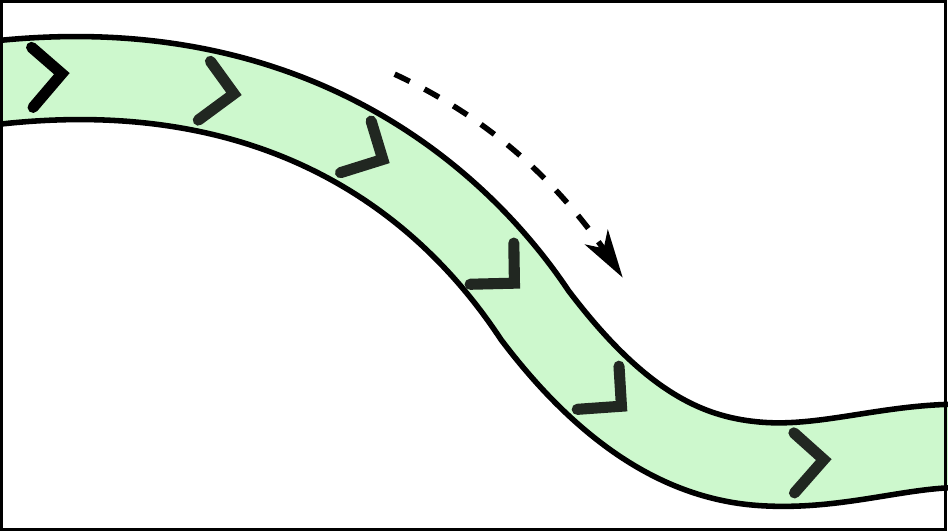}}
    \label{fig:arrow-glyphs}
  }
	 \subfigure[Transparency gradient]{
    \resizebox{0.30\columnwidth}{!}{\includegraphics{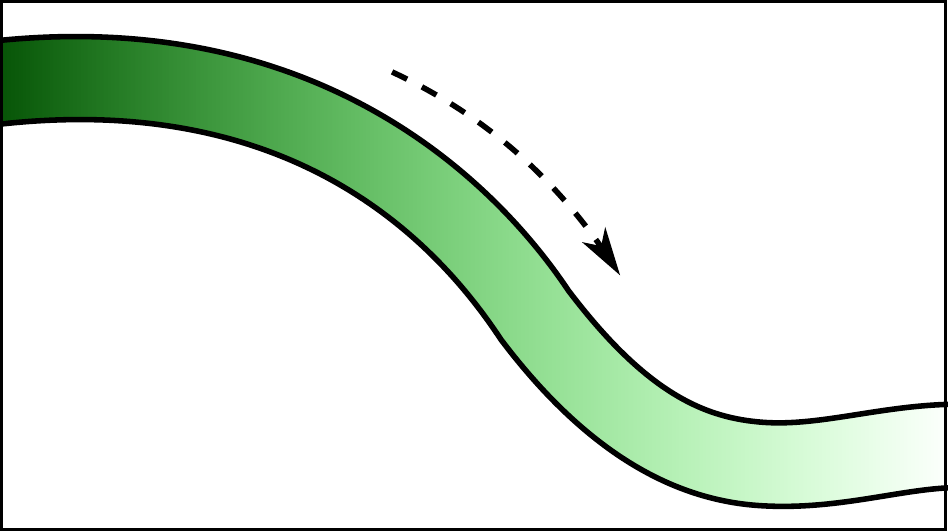}}
    \label{fig:trans-gradient}
  }
  \subfigure[Tapered line]{
    \resizebox{0.30\columnwidth}{!}{\includegraphics{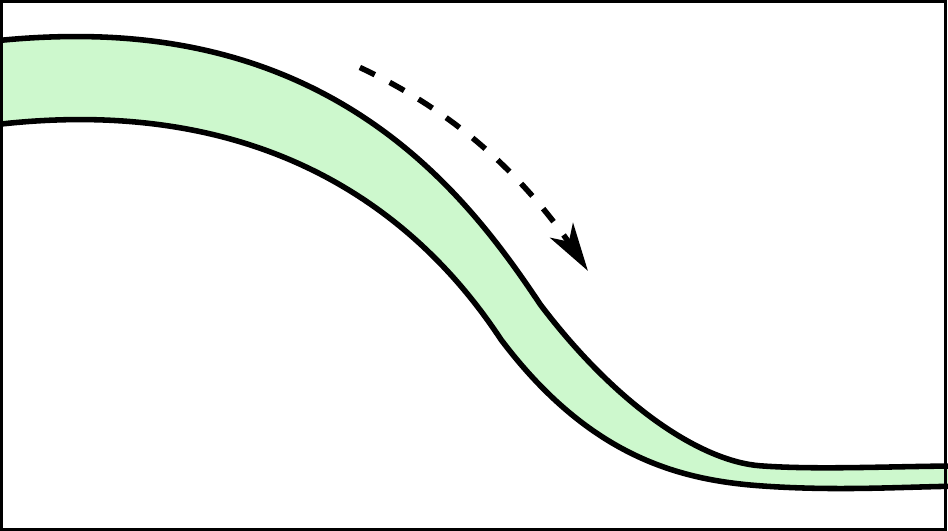}}
    \label{fig:tapered-line}
  }
  \caption{Three visual design alternatives for conveying route
    directionality. 
		}
  \label{fig:direction}
\end{figure}

\textbf{Route Direction:} 
Visualizing route direction (R3) is challenging due to limited space and potential for visual clutter.
Inspired by Holten et al.~\cite{Holten2011, Holten2009b}, we propose
three alternative designs for conveying direction:

\vspace{-0.3em}
\setlist{leftmargin = 3mm}
\begin{itemize}
\itemsep0em
\item\textbf{Arrow glyph:} An arrow is an intuitive
  representation to encode direction.
  We place multiple arrow
  glyphs at regular intervals along the entire route to \textcolor{red}{fulfill} R3 (Figure~\ref{fig:arrow-glyphs}).
  The size of the arrow is scaled by the width of the route line to
  save space and reduce clutter.

\item\textbf{Transparency gradient:}
    This method uses a gradient of increasing transparency to convey the directionality (Figure~\ref{fig:trans-gradient}).
  The gradient can be applied to an entire route, or repeated for
  each leg.
  
\item\textbf{Tapered line:} This design uses a variable-width
  route line that starts out at full thickness on the start node and
  tapers to smaller thickness at the end
  (Figure~\ref{fig:tapered-line}).
  As for the transparency gradient, this design can be applied
  globally for an entire route, or repeated over and over again for
  each leg on the route.
\end{itemize}

While we considered using a color scale such as the red-to-green
mapping proposed by Holten et al.~\cite{Holten2011, Holten2009b}, we
ultimately decided against this visual design because we already use
categorical color to differentiate routes.

\subsection{Visual Design: Nodes}\label{sec:node-design}
\begin{figure}[htb]
  \centering
  \subfigure[Concentric rings]{
    \resizebox{0.3\columnwidth}{!}{\includegraphics{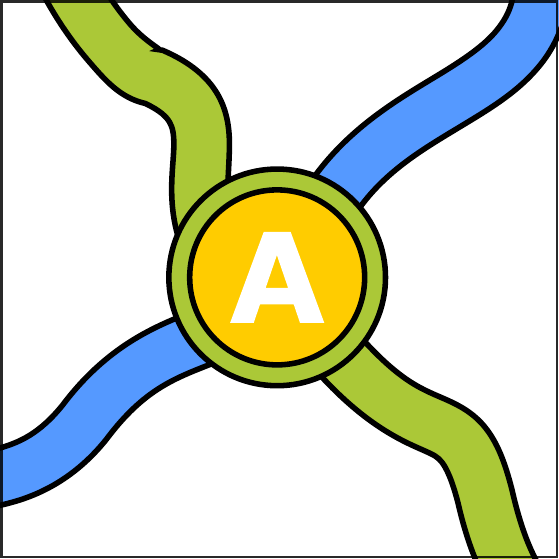}}
    \label{fig:concentric-rings}
  }
  \subfigure[Cookie bites]{
    \resizebox{0.3\columnwidth}{!}{\includegraphics{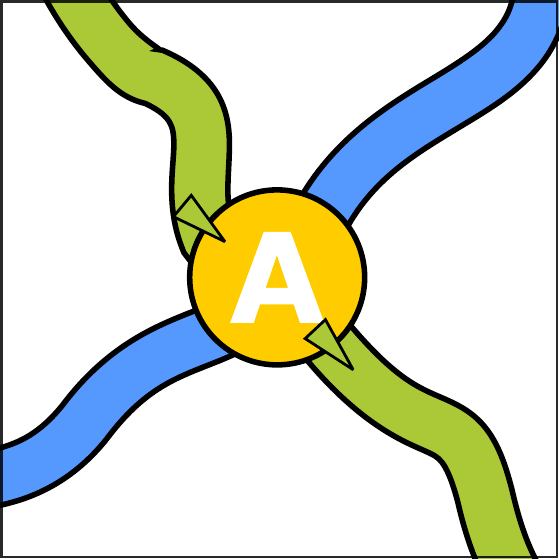}}
    \label{fig:cookie-bites}
  }
  \subfigure[Integrated arrows]{
    \resizebox{0.3\columnwidth}{!}{\includegraphics{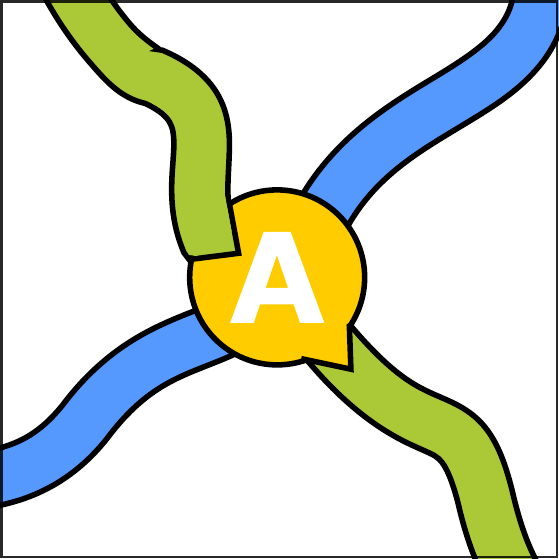}}
    \label{fig:integrated-arrows}
  }
  \caption{Design alternatives for nodes where the blue route (SW
    $\rightarrow$ NE) passes by a node, but where the green route (NW
    $\rightarrow$ SE) makes a stop.}
  \label{fig:node-design}
  \vspace{-0.5em}
\end{figure}

Multiple routes may share the same locations as starting or stopping points; for example, in
a supply-chain network, all routes tend to begin and end at a few
distribution centers.
This poses a challenge to effectively identifying and distinguishing
those points of multiple routes (R2).
For this reason, we propose three visual designs for
communicating this information:

\vspace{-0.3em}
\setlist{leftmargin = 3mm}
\begin{itemize}
\itemsep0em
\item\textbf{Concentric rings:}
  Inspired by metro maps~\cite{Degani2013}, we overlay concentric rings with different colors on the node to indicate whether it is a waypoint for the corresponding color-coded route (Figure~\ref{fig:concentric-rings}).

\item\textbf{Cookie bites:} 
  Here, arrows pointing into the center of
  the node convey a stopping point; an arrow pointing out conveys the
  route leaving the node (Figure~\ref{fig:cookie-bites}).

\item\textbf{Integrated arrows:}
  Similar to cookie bites, but the arrowheads are integrated into the route lines themselves to reduce visual clutter (Figure~\ref{fig:integrated-arrows}).
\end{itemize}

\vspace{-1.2em}

Our current implementation uses concentric rings because of
their ubiquity in metro and subway maps.

\begin{figure*}[thb]
  \centering
  \subfigure[Linear KDE.]{
    \resizebox{0.3\textwidth}{!}{\includegraphics{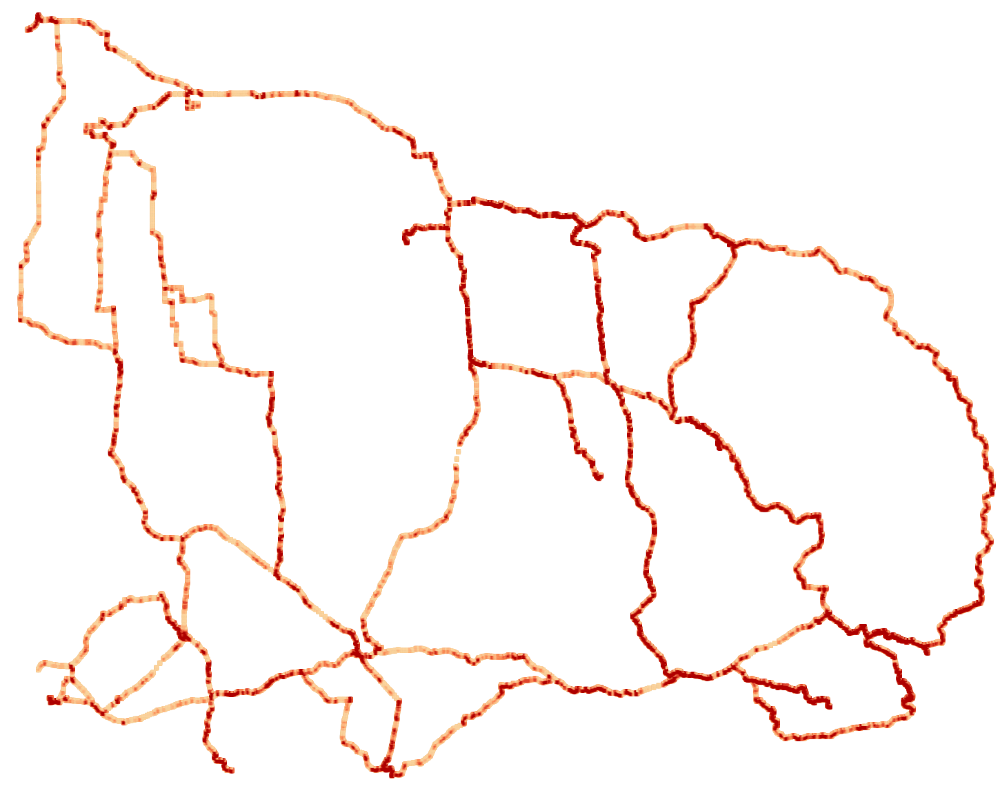}}
    \label{fig:kde}
  }
  \subfigure[Thinning.]{
    \resizebox{0.3\textwidth}{!}{\includegraphics{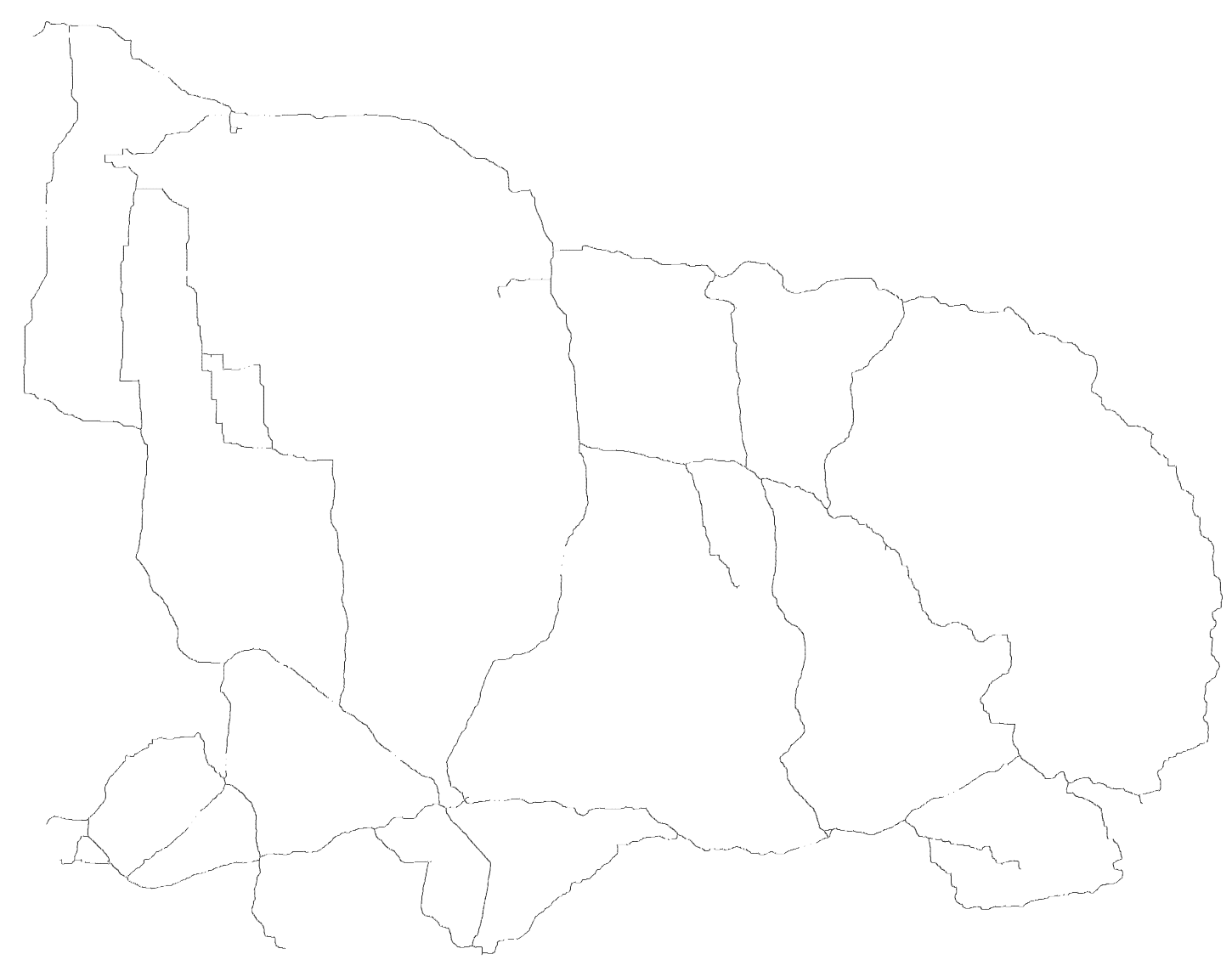}}
    \label{fig:thinning}
  }
  \subfigure[Bifurcation Points]{
    \resizebox{0.3\textwidth}{!}{\includegraphics{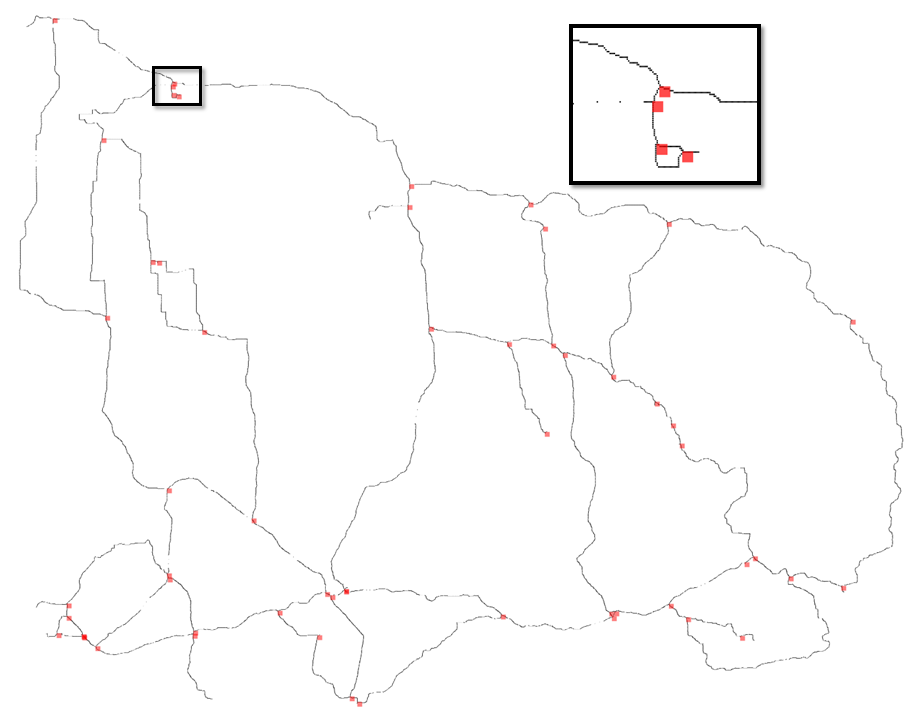}}
    \label{fig:bifurcation}
  }
  \caption{An illustration of applying linear KDE in (a) and then thinning algorithm to extract the skeleton of the road network in (b). In (c), red rectangles highlight the extracted results of bifurcation points.}
  \label{fig:pre-process}
  \vspace{-0.5cm}
\end{figure*}

\subsection{Implementation}\label{sec:impl}
Our route packing system is implemented using the web-based D3~\cite{Bostock2011}  toolkit.
Map and route network data are retrieved using the Bing Maps REST
services\footnote{\label{note:bingmaps}\url{https://msdn.microsoft.com/en-us/library/ff701713.aspx}}.
For each route, we query for the shortest driving distance and retrieve high-precision coordinates along the routes.

Bing Maps's outputs include thousands of sequential pairs of latitude and longitude. To simplify the raw graph data model, we first detect all overlapping routes (i.e., e4, e5, e6, e7 between leg (B, F) in Figure~\ref{fig:route-network}). For each shared route, we extracted its start point (i.e., B in Figure~\ref{fig:route-network}) and end point (i.e., vertice between e7 and e8 in Figure~\ref{fig:route-network}) as the crucial vertices to build a pruned graph. After that, we juxtapose shared routes side by side and minimize the number of crossings between routes (Algorithm 1). We preserve the relative shared routes displacement order within an edge between two crucial vertices and rearrange the displacement order for its consecutive edges if needed. In what follows, we elaborate on the details of our approach. 


\subsubsection{Detecting Shared Routes}
We first detect shared sections of different routes, which may be occluded and hinder the effective recognition of routes (Figure~\ref{fig:packing}).
We use an approach based on linear KDE adopted from Lampe and Hauser~\cite{Lampe2011} to extract the skeleton of the road network and identify groupings based on proximity.
Figure~\ref{fig:kde} shows the result of applying linear KDE on the raw output from Bing Maps APIs.
After converting to a binary image, we then apply the Zhang-Suen thinning algorithm to extract a skeleton of the route network~\cite{zhang1984} (Figure~\ref{fig:thinning}). 
Finally, we apply
a bifurcation detection algorithm~\cite{mehtre1993} to identify intersections (Figure~\ref{fig:bifurcation}) as crucial vertices to build a pruned graph.

\renewcommand{\figurename}{Algorithm}
\setcounter{figure}{0}
\begin{figure}[htb]
  \raggedright
  \resizebox{0.92\columnwidth}{!}{\includegraphics{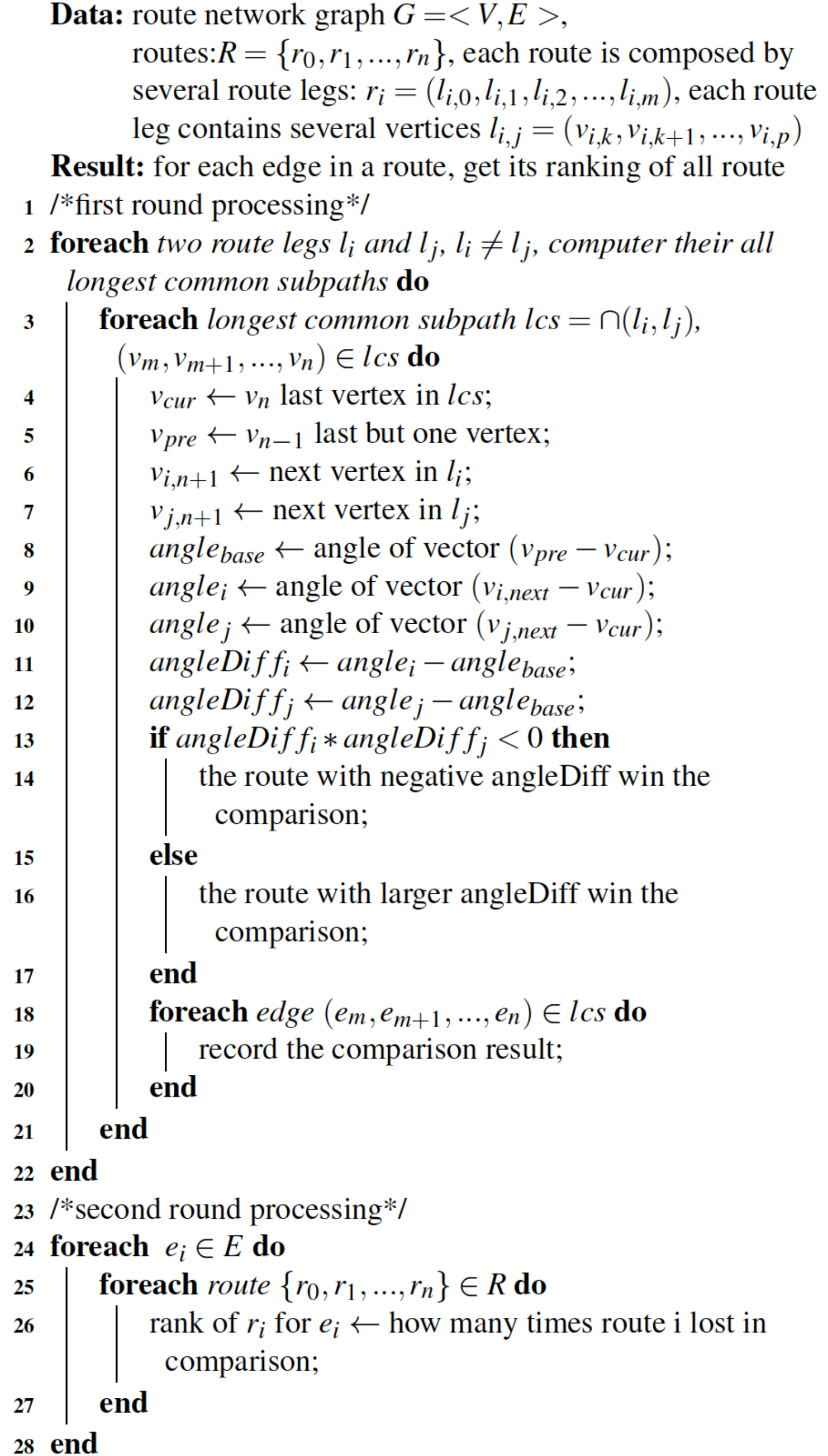}}
  \caption{Minimize route crossings algorithm.}
  \label{alg:rank}
	\vspace{-0.5cm}
\end{figure}
\renewcommand{\figurename}{Figure}
\setcounter{figure}{6}

\subsubsection{Shifting and Packing \textcolor{red}{Overlapping} Edges}
After detecting shared routes, we obtain a graph of vertices (intersections and destinations) and intermediary connecting edges. 
Each vertex in the graph is assigned a unique integer ID and each route leg as a sequence of vertices with these IDs. 
Two phases of processing are applied to obtain the proper ranking and the displacement of shared edges. 
\textcolor{red}{To minimize crossings and reduce visual clutter}, we adopted the MLCM algorithm in~\cite{Nollenburg2010} and customized it for \textit{Route Packing} to account for turns greater than 90 degrees. 

We consider each vertex \textcolor{red}{as} an intersection of traffic either merging or diverging, and based on that, we \textcolor{red}{determine} the displacement order of different edges. 
This strategy has drawbacks when we draw an entire route along multiple edges, e.g. a large shift of displacement for two consecutive edges may generate extra curves. 
Our displacement algorithm \textcolor{red}{can reduce} such turbulence and minimizes crossings along the direction of a route: Algorithm 1 takes the reconstructed route network graph as input and generates the relative rank of each route. We generalize the comparison between two routes as diverging traffic.

\begin{figure}[htb]
  \centering
  \resizebox{0.5\columnwidth}{!}{\includegraphics{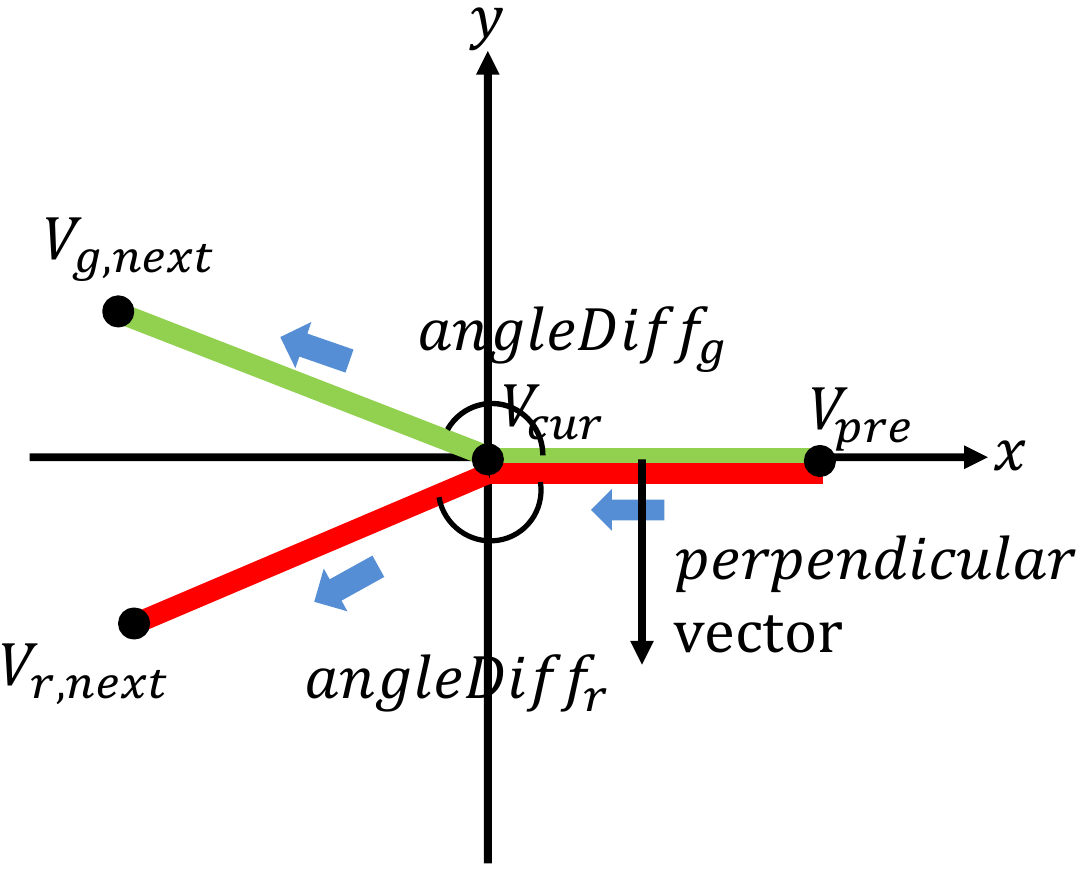}}
  \caption{A generalized scenario of calculating the order of routes for a shared subpath.}
  \label{fig:direct_compare}
  \vspace{-0.5em}
\end{figure}


Suppose two route legs share at least one edge.
The displacement of routes on shared edges is decided by the angles where the routes leave the overlap subpath.
Figure~\ref{fig:direct_compare} demonstrates a visual example corresponding to pseudocode line 4--17 in Algorithm 1.
First, we need to identify the longest common subpaths between all route legs. The displacement among the common subpath does not change, until two routes separate. 
As shown in Figure~\ref{fig:direct_compare}, we put the last vertex in shared subpath on the center of the coordinate system. 
Then we align the second to last vertex in the shared subpath on the positive x Axis by subtracting the angle of vector $(V_{pre} - V_{cur})$. 
The range of an angles is $(-180,180]$. 
The perpendicular vector is defined as counter-clockwise 90 degrees to the vector $(V_{cur} - V_{pre})$ along the moving direction.  
The red route tends to be top towards the direction of the perpendicular vector. 
After comparison of all shared subpaths, the rankings of all edges are determined. 
The ranking of a route for an edge towards its perpendicular vector direction \textcolor{red}{is equal to} how many time it lost in comparisons.

\begin{figure*}[thb]
  \centering
	\subfigure[Arrow glyph (AG).]{
    \resizebox{0.3\textwidth}{!}{\includegraphics{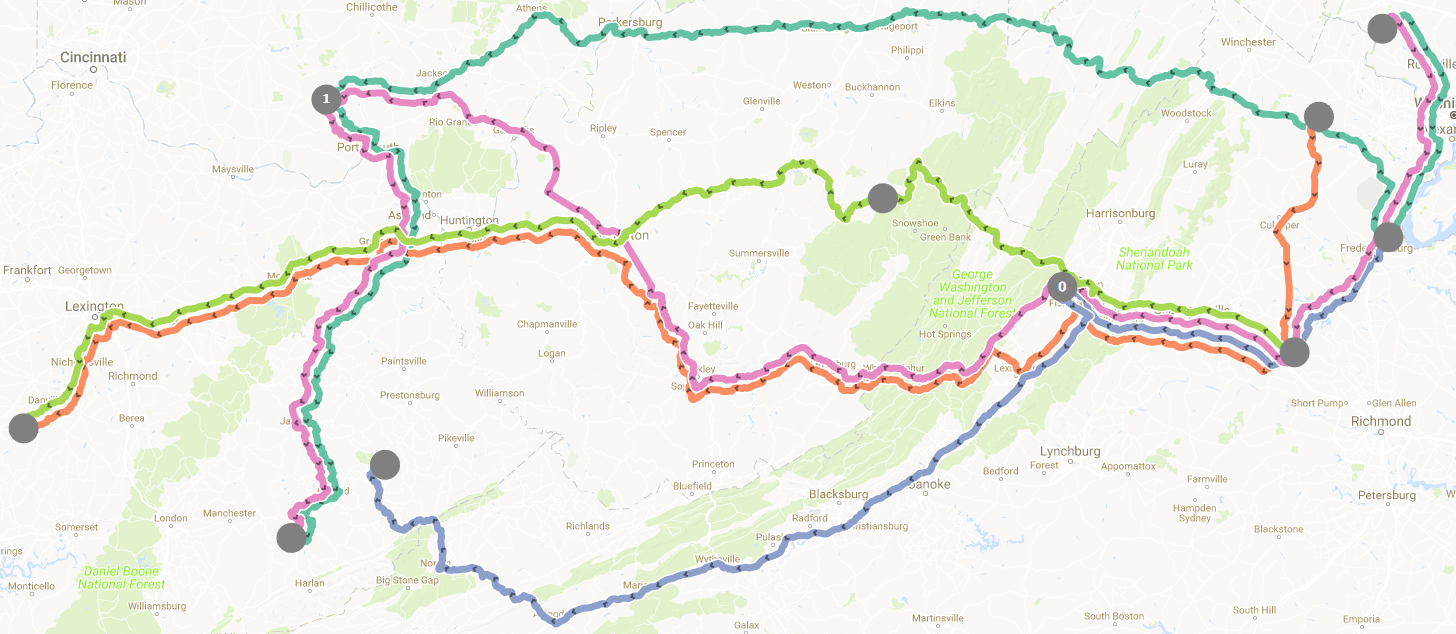}}
    \label{fig:64_AG_cut}
  }
  \subfigure[Transparency gradient (TR).]{
    \resizebox{0.3\textwidth}{!}{\includegraphics{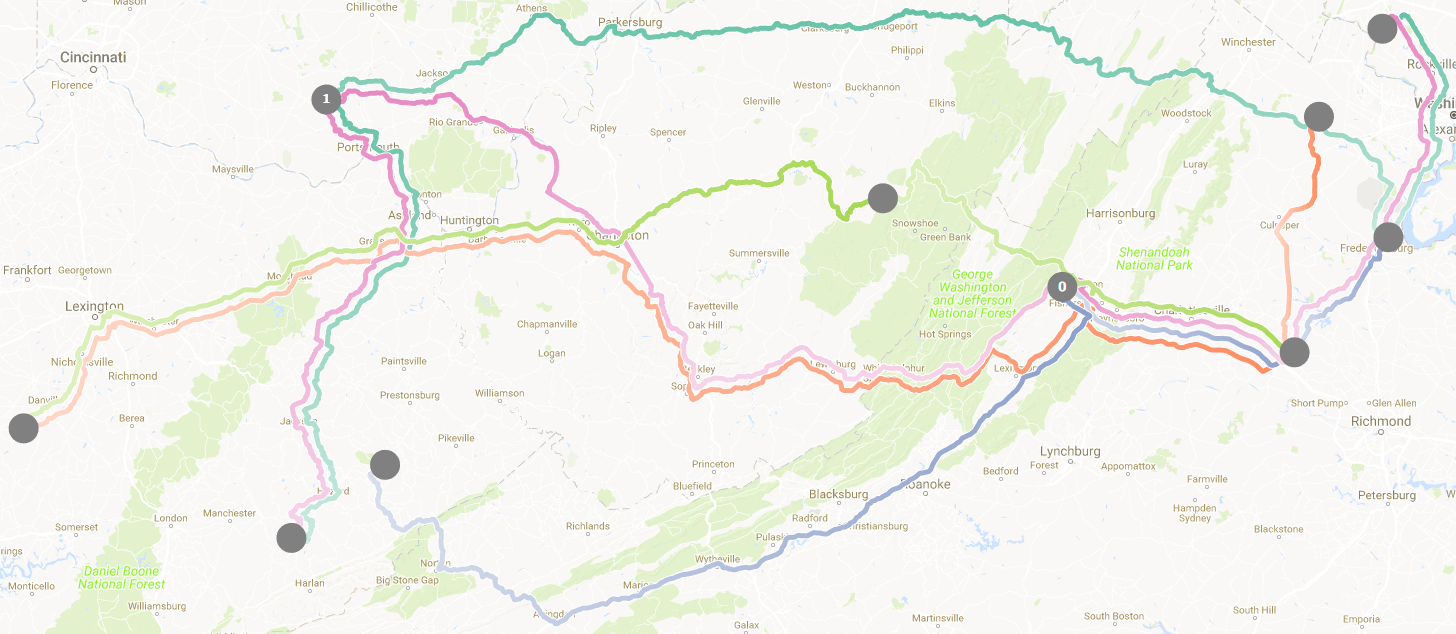}}
    \label{fig:64_Tr_cut}
  }
  \subfigure[Local tapered line (LT).]{
    \resizebox{0.3\textwidth}{!}{\includegraphics{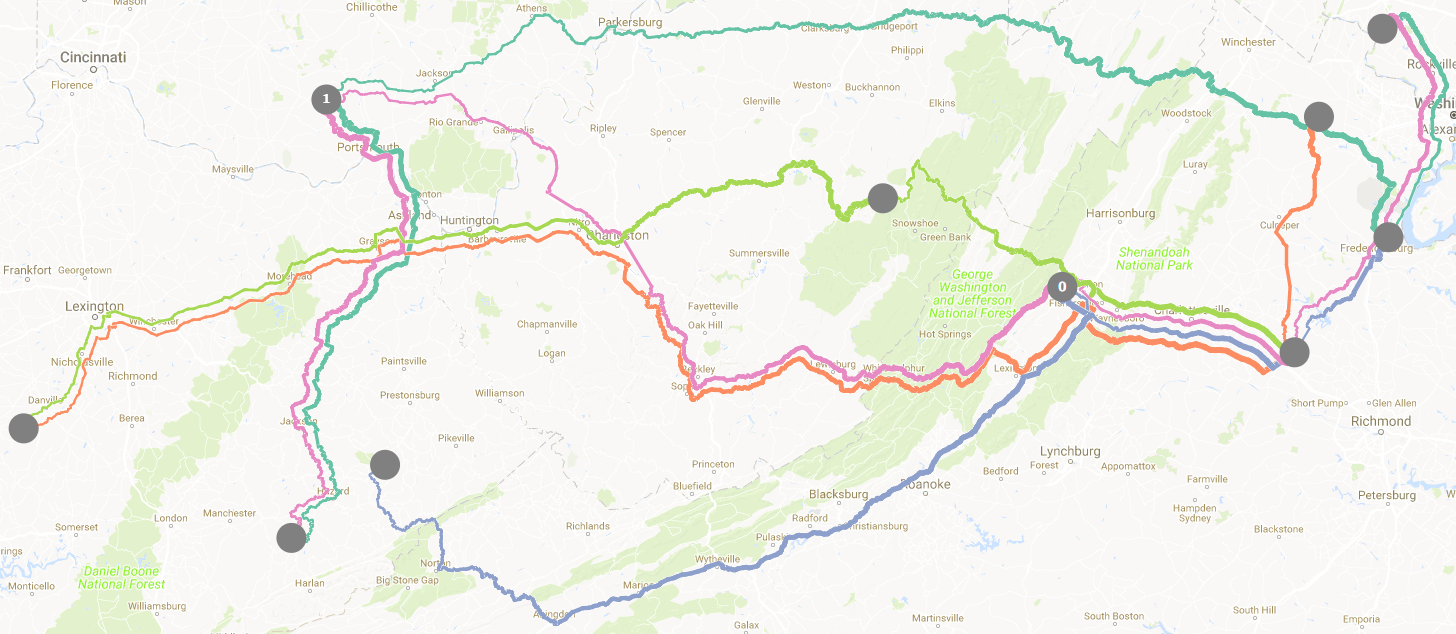}}
    \label{fig:64_LT_cut}
  }
  \caption{Three variants of route packing using different
    directionality methods.}
    \vspace{-0.5cm}
  \label{fig:all-designs}
\end{figure*}

\subsubsection{Transparency Gradient and Tapered Line}
We use a gradient of increasing transparency to indicate route direction, where the transparency 
increases along the route from the starting point until the end point.
For the current implementation, we set 35\% and 100\% as the minimum and maximum opacity.
Similarly, for the tapered line, we utilize a variable-width route line
that begins at maximum thickness at the start node, and tapers down to
a minimum thickness by the end node.
The width value is chosen to avoid unnecessary space usage while
facilitating visual identification of the change in line width.
In our implementation, we select 2 and 6 pixels as the minimum and
maximum width.

\section{User Study}\label{sec:user_study}
We conducted a crowdsourced user study using Amazon Mechanical
Turk
to validate our design choices and determine the
best directionality encoding for route packing
(Table~\ref{table:exp_set}).
Our participant pool were all crowdworkers. 
Route maps are common in daily life, and thus, it makes sense to engage a general population in the study.
We recruited 132 (56 female) participants using Amazon Mechanical Turk
with an age range between 20 to 68.
Color blind users were excluded through self-reporting.
Participants were paid \$0.80 and spent an average of 10 minutes on the experiment.
Sessions were performed using a web browser on desktops, laptops, and a few tablets.

\begin{table}[ht]
  \caption{Experiment setting.}
  \label{table:exp_set}
  \scriptsize
  \begin{center}
    \begin{tabular}{c | c | c}
      \hline
      \textbf{Group ID} & \textbf{Representations} & \textbf{Participants} \\
      \hline\hline
      \multirow{2}{*}{Group 1} & Transparency Gradient, Arrow Glyph (TRA) & \multirow{2}{*}{32} \\
      \hhline{} & Local Tapered Line, Arrow Glyph (LTA) &\\
      \hline
      \multirow{2}{*}{Group 2} & Transparency Gradient (TR) & \multirow{2}{*}{35} \\
      \hhline{} & Local Tapered Line (LT) & \\
      \hline
      \multirow{2}{*}{Group 3} & Arrow Glyph (AG) & \multirow{2}{*}{37} \\
      \hhline{} & Global Tapered Line, Arrow Glyph (GTA) & \\
      \hline
      \multirow{2}{*}{Group 4} & Global Tapered Line (GT) & \multirow{2}{*}{28} \\
      \hhline{} & Arrow Glyph, Concentric Ring (AGR) & \\
      \hline
    \end{tabular}
  \end{center}
  \vspace{-1.5em}
\end{table}

\subsection{Dataset and Task}
Our experimental trials were designed by choosing different locations
in the U.S.\ to generate random georeferenced graphs.
To generate a georeferenced graph, we first randomly selected 10 nodes
for a specified geospatial bounding box.
To ensure similar task complexity, the 10 nodes were
connected in an undirected graph.
From these 10~nodes, we generated a total of 3 to 5 routes by randomly selecting 3~to~5 nodes (i.e., stops) to generate a connected route.
We then retrieved a georeferenced subgraph for each route using the
Bing Maps REST services API.
Finally, we applied route packing on the resulting graph.



In the experiment, participants were shown a map visualization with
two highlighted nodes.
They were asked to determine whether the two nodes (i.e., stops) were
connected through a route; that is, whether there existed a route that
allowed them to travel from a given source to destination. 
We chose this task to ensure that our technique is able to visualize a route between two nodes effectively, even when multiple routes (sharing the same legs or crossing) are present. 
Jumping from one route to another route to reach the
destination was explicitly forbidden, but there could be intermediate
stops along the path between A and B, as shown in Figure~\ref{fig:directvssteps_1_cp}.
We emphasized the importance of directionality in the route network,
but did not limit the number of intermediate nodes between stop A and
B.

\subsection{Experimental Factors and Design}
\label{userStudy:experimentalFactors}
\begin{figure}[h]
  \centering
  \subfigure[]{
    \resizebox{0.45\columnwidth}{!}{\includegraphics{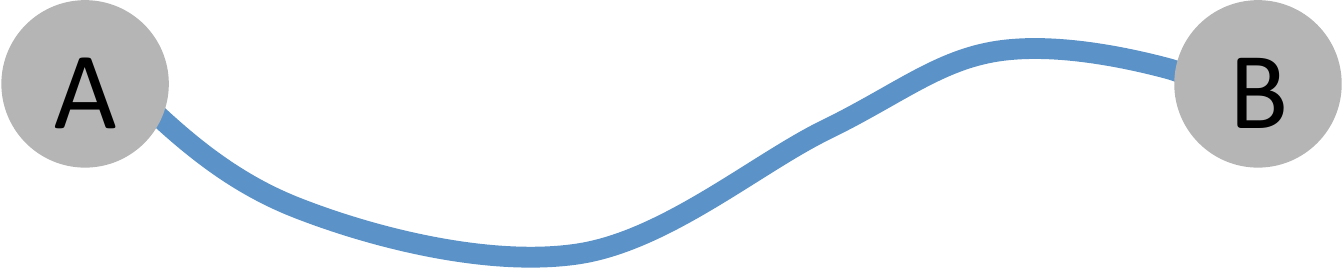}}
    \label{fig:directvssteps_1_cp}
  }
  \subfigure[]{
    \resizebox{0.45\columnwidth}{!}{\includegraphics{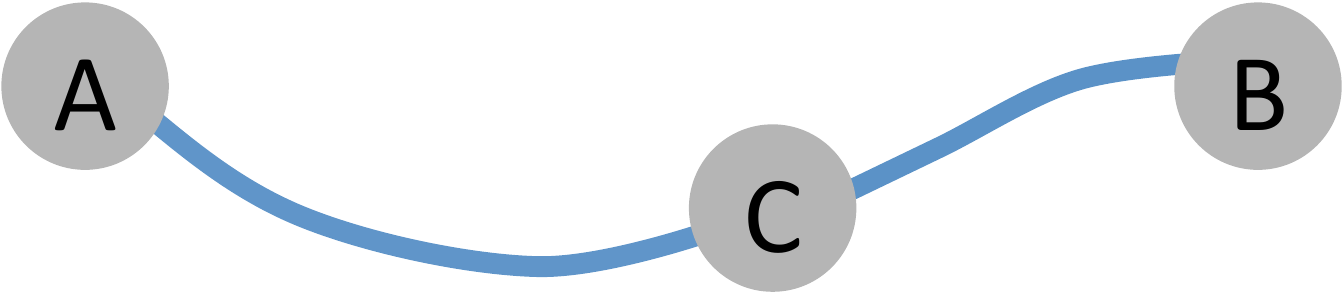}}
    \label{fig:directvssteps_2_cp}
  }
  \caption{Complexity levels of user study questions.}
  \label{fig:directvssteps}
  \vspace{-0.5em}
\end{figure}

The two experimental factors tested in the experiment are visual representations to convey directionality and levels of complexity. It is important to note that these two factors were tested in the context of multiple routes, and therefore, testing route packing in general. 
Firstly, We have three major designs for preserving route directionality: arrow
glyph (Figure~\ref{fig:64_AG_cut}), transparency gradient (Figure~\ref{fig:64_Tr_cut}), and tapered line (Figure~\ref{fig:64_LT_cut}).
The tapered line can be applied at two different scales: localized
tapered line (stop to stop), and global tapered line (an entire
route).
Three different visual encoding strategies for indicating directionality were also evaluated: transparency gradient, local tapered line, and global tapered line.
All these strategies can be applied alone or in combination with
arrows that results in dual encoding of directionality.
Beyond connectivity, we also studied the utility of the concentric
ring design (Section~\ref{sec:node-design}).
Table~\ref{table:exp_set} shows the final set of 
representations we tested in the experiment. 
Secondly, the two complexity levels of questions tested in our experiment are: (a) stop A is directly connected with stop B, or (b) stop A is connected with stop B through an intermediate node (Figure~\ref{fig:directvssteps}).
This connectivity task is equivalent to that of Holten et
al.~\cite{Holten2011, Holten2009b}.

\vspace{-.5em}


To minimize the number of different visual representations for each
participant and to keep each session under 10 minutes, we conducted
the experiment in four groups.
For each group, the participants were presented with two visual
representations ($2 \times Vis$), and were given ten trials per
representation: $5 \times two-stop$ (directly connected) routes, and
$5 \times three-stop$ (one intermediate node) routes.
This total of twenty trials in each group was generated using the
same georeferenced graph data, but with different representation
methods.
We measured both accuracy and time spent on each trial.

After conducting the user study with 135 number of participants, we
analyzed the correctness and completion times for different visual
representations by averaging the results of all trials for the
specific representation. 
We removed the data of 29 participants from our analysis as their performance was determined to be similar to chance (i.e., their results indicated that they had randomly clicked on any answer). 
\vspace{-0.5em}
\begin{table}[htb]
  \caption{Effects of factors on completion time.}
  \label{table:exp_effect}
  \scriptsize
  \begin{center}
    \begin{tabular}{c | c | c | c | c}
      \hline
      \textbf{Effect} & \textbf{Num OF} & \textbf{Den DF} & \textbf{F Value} & \textbf{Pr \textgreater F} \\
      \hline\hline
      \textbf{Vis} & 7 & 2216 & 6.88 & \textless.0001 \\
      \hline
      \textbf{Difficulty} & 1 & 2216 & 1.74 & 0.1867 \\
      \hline
      \textbf{Vis*Difficulty} & 7 & 2216 & 2.00 & 0.0519 \\
      \hline
    \end{tabular}
  \end{center}
  \vspace{-1.5em}
\end{table}

\begin{figure}[htb]
  \centering
  \subfigure[Average correctness.]{
    \resizebox{0.47\columnwidth}{!}{\includegraphics{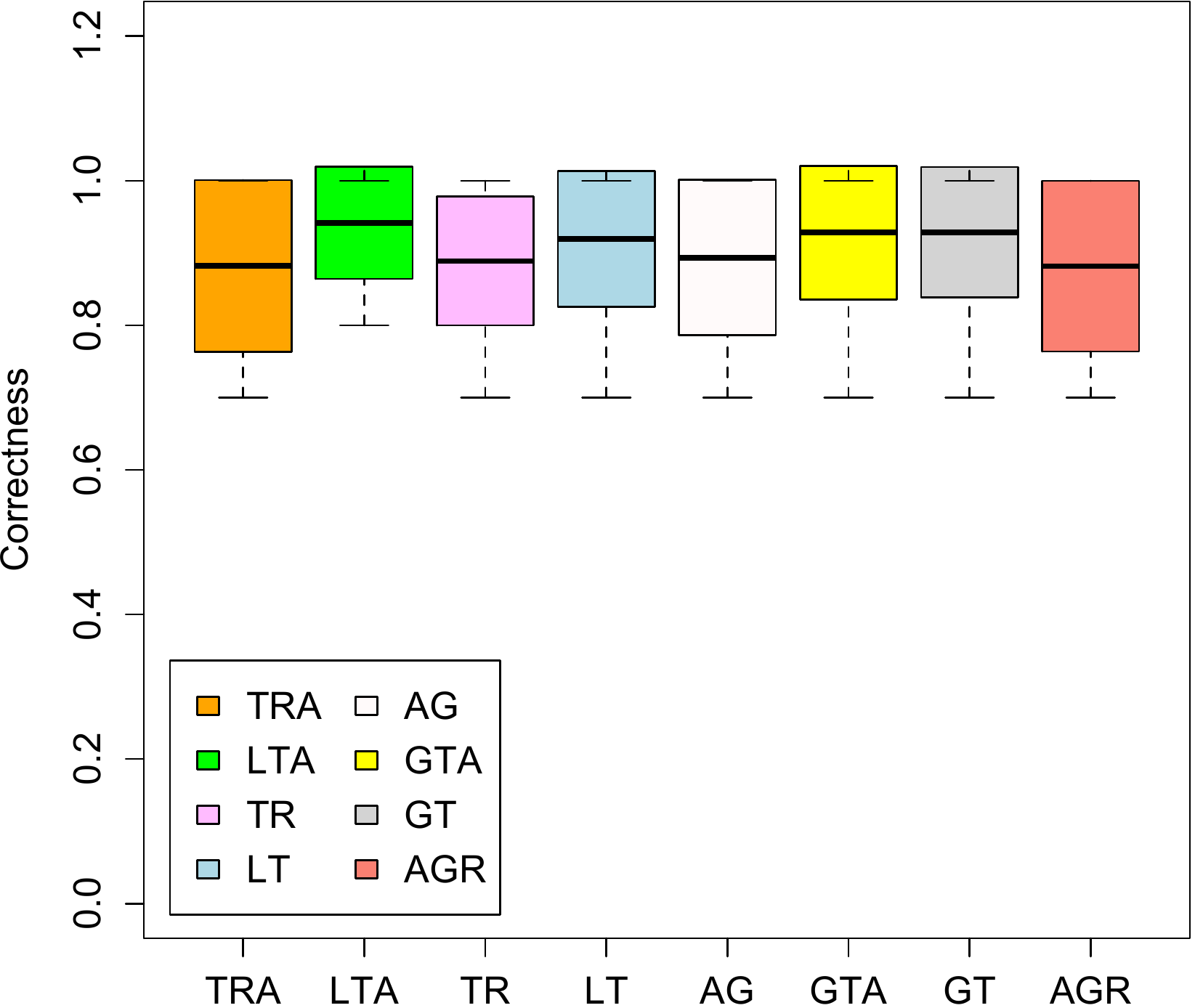}}
    \label{fig:correctness}
  }
  \subfigure[Average completion time.]{
    \resizebox{0.47\columnwidth}{!}{\includegraphics{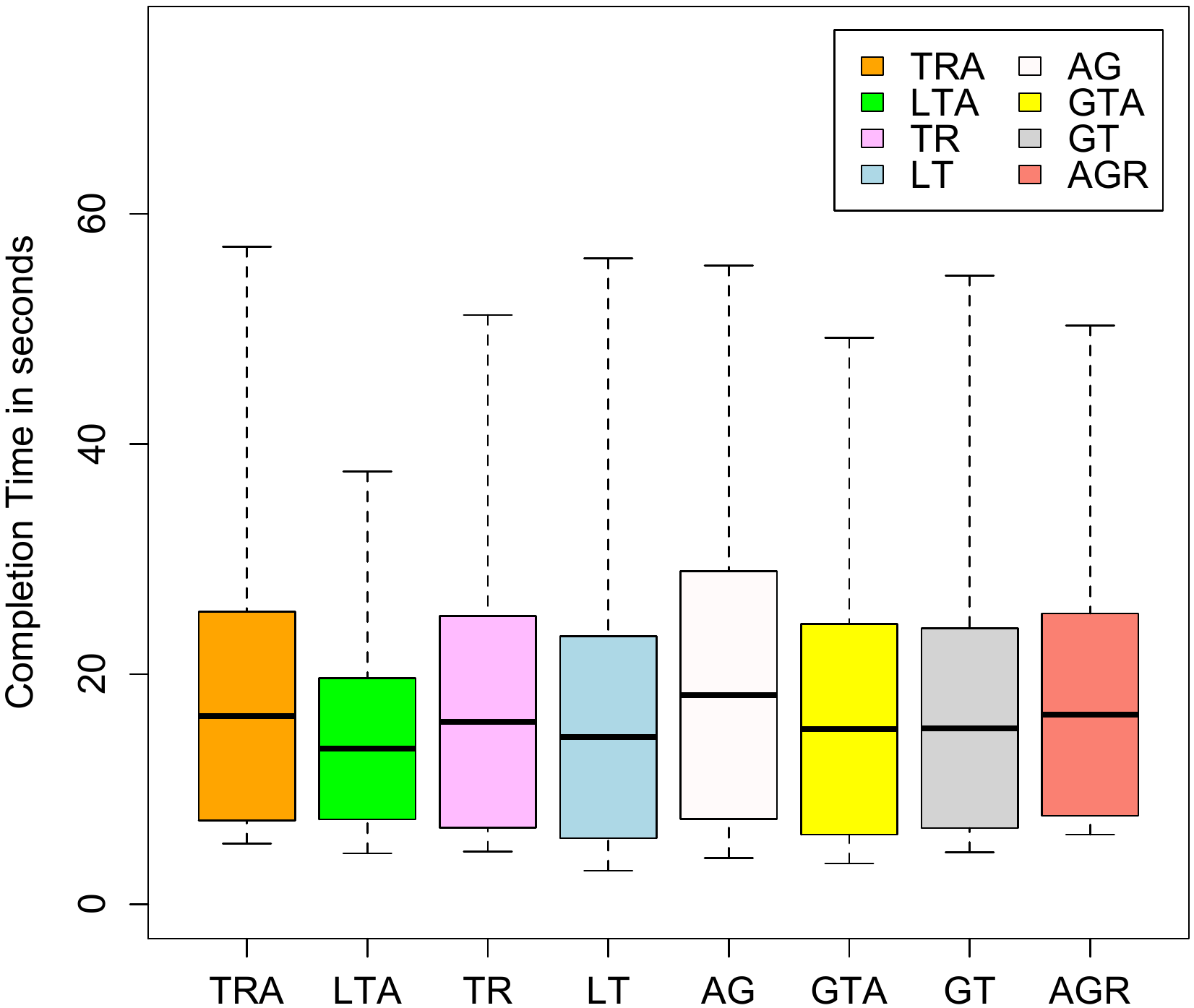}}
    \label{fig:completiontime}
  }
  \caption{Correctness and completion time plots for eight route
    visual representations.}
  \label{fig:result}
	\vspace{-1.2em}
\end{figure}


\subsection{Experiment Results}
\vspace{-0.3em}
Figure~\ref{fig:correctness} summarizes the correctness results for
the eight different visual representations.
The vertical axis shows the correctness ratio scaled from 0 to 1.
We conducted a logistic regression with binomial distribution on the
correctness ratings, and found no significant effect of visualization
on correctness: $F(1, 7) = 0.46, p = 0.8650$.
This indicates that the different visual representations have similar
correctness results (i.e., the different representations convey the
information with similar accuracies).

We measured the completion time for each representation by averaging
across all trials. 
We analyzed the results using a repeated-measures analysis of variance
(assumptions met).
Visual representation had a significant effect on completion time:
$F(1, 7) = 6.88, p < .001$ (Table~\ref{table:exp_effect}).
Figure~\ref{fig:completiontime} shows average completion time as a
function of the eight route visual representations.
These two completion time results indicate that the different route
visual representations have significant effect on completion time in
our study. 


The participants completed a post-test questionnaire for providing
subjective feedback and overall preference on visualizations.
76 participants ($\approx$57.5\%) preferred arrow glyphs (AG), followed
by transparency (33 participants$\approx$25.0\%), and tapered design
(23 participants$\approx$17.5\%).
It is notable that participant preference is different from their
performance.
Participants achieved the highest speed with the tapered design
(tapered$<$TR$<$AG).
We compared the preference of individual designs to actual completion time.
The participants who preferred AG spent 20.2 seconds on average, while the participants who preferred the tapered design spent 14.7 seconds on average.
To sum up, AG might be easier to understand and more preferable due to familiarity, but the tapered design is a better choice in actual tasks for following directions of routes.

\begin{figure}[htb]
  \centering
  \resizebox{0.8\columnwidth}{!}{\includegraphics{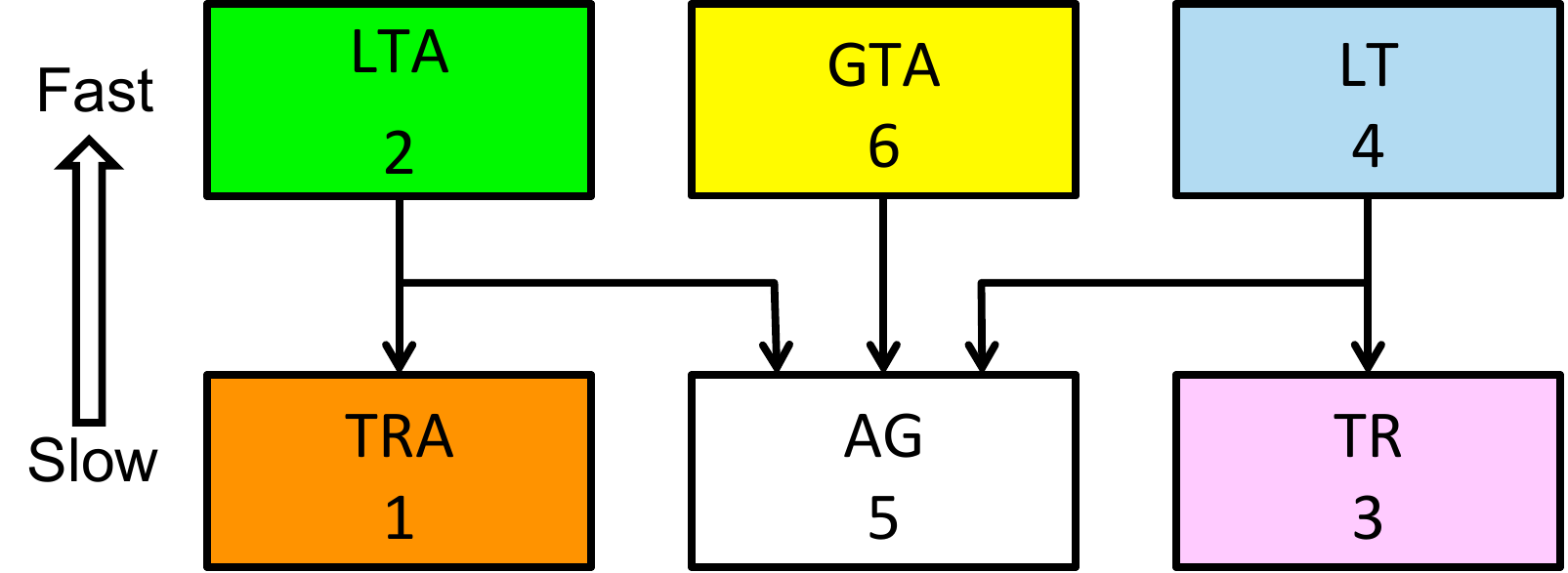}}
  \caption{Pairwise relationships of visualization by completion
    time (Tukey HSD, $p < 0.05$).}
  \label{fig:study_vis_relationship}
    \vspace{-0.55cm}
\end{figure}


We summarize our findings from the user study as follows
(Figure~\ref{fig:study_vis_relationship}):

\vspace{-0.3em}
\setlist{leftmargin = 3mm}
\begin{itemize}
\itemsep0em
\item Users spent less time solving trials using local tapered lines with arrows (LTA) than using transparency gradient with arrows (TRA) and arrow glyphs (AG).
\item Users spent less time solving trials using local tapered lines (LT) than using transparency gradient (TR) and arrow glyphs (AG).
\item Users spent less time solving trials using global tapered lines with arrows (GTA) than with using arrow glyphs (AG).
\item There were no significant differences between arrow glyphs with concentric rings (AGR) and other visualizations.
\item Overall, tapered representation performed the best.
\end{itemize}
\vspace{-0.5em}

\subsection{Explaining the Results}
\vspace{-0.5em}
The results from our study validate our design objectives that different route visual representations would garner different performances. 
We also find that design add-ons (arrows, concentric rings) do not
have a major effect on completion time. 
These results provide guidance for \textcolor{red}{the optimal} design of route
visualizations.
 
In Figure~\ref{fig:study_vis_relationship}, we demonstrate the
relationships between the 6 different route visual representations
based on our posthoc analysis
(significant to $p<0.05$).
We find that tapered line design \textcolor{red}{is} the best design compared to the other \textcolor{red}{options} because all of the techniques involving tapered designs---GRA,
LTA, and LT---are superior to some other techniques even if there is no global order.
More specifically, both LTA and LT have a shorter completion time compared to TRA, TR, and AG. 
In addition, GTA performed better than AG. 
We conclude that the tapered line design outperforms the other
visualizations in conveying route directionality.
This result is also consistent with the work by Holten et
al.~\cite{Holten2011, Holten2009b}, which found that a tapered design was optimal for visualizing directed edges in graphs.
However, we find no significant difference in performance 
between the transparency design and routes with arrows. In addition, add-on arrows do not increase user performance with any of the types of visual design. Surprisingly, our analysis shows that there is no advantage to adding concentric
rings at stops for conveying waypoints. 
\textcolor{red}{A case study that shows the utilization of a tapered design and concentric rings for flight planning in the aviation domain is provided in the appendix\footnote{\url{https://docs.lib.purdue.edu/purvacsup/1}}.}


\subsection{Generalizing the Results}
\vspace{-0.5em}
\textcolor{red}{Our design is applicable to any closed route network for the visualization of 10-20 routes simultaneously. 
The number of routes is limited because the route packing technique consumes extra spaces to juxtapose routes while preserving their geographic information.}

Furthermore, we conducted our study using Mechanical Turk.
Our participants were diverse with different proficiency levels.
This indicates that our participant pool is fairly representative of the general population.
We also obtained consistent results from our participants,
which is another indication that our findings generalize to other audiences.
\textcolor{red}{Though our subjects were non-experts, we expect that our results will hold with route planning experts.}

Finally, the scalability of our techniques is difficult to assess given our evaluation results.
The tasks were relatively small in scope;
see Figures
\ref{fig:all-designs} for example images.
\textcolor{red}{To evaluate the effectiveness of conveying directionality,} we only included 10 nodes (i.e., potential stops) in our route network trials.
For larger and more dense route networks, route packing may cause too many routes to be packed in parallel, thereby consuming too much screen space. 
For instance, the packing strategy would fail if a dozen routes share the same tight U-turn on the map.
\textcolor{red}{Thus, additional filtering and zooming interactions need to be incorporated to support the analysis of larger networks.}

\vspace{-0.3em}
\section{Conclusion and Future Work}\label{sec:conclusion}
\vspace{-0.5em}
We presented a novel technique for visualizing routes on geographic maps called \textit{route packing}, where the focus is to remain faithful to the identity, directionality, and
weight of each individual route.
To validate the utility of the technique, we also presented results
from a controlled user study.
The results point to the overall utility of the technique as well as
the tradeoffs involved in the complexity and detail of the dataset to be visualized.
\textcolor{red}{In future research, we plan to apply route packing to a visual analytics
tool for supply-chain logistics and simulation, as well as exploring its potential in other domains.} 
Given that our current study in this paper was conducted with
participants drawn from a general population, we are interested in
determining how domain experts in supply-chain
management \textcolor{red}{may utilize our visualization. A comparative experiment against other techniques tackling route overlap is another possible direction that future research can undertake.}

\vspace{-0.3em}
\section{Acknowledgments}
\vspace{-0.5em}
This work is funded in part by the U.S. Department of Homeland Security VACCINE Center under Award Number 2009-ST-061-CI0003.


\bibliographystyle{ieeetr}
\bibliography{routevis}

\end{document}